\documentclass[10pt, twocolumn]{IEEEtran}
\usepackage{bbding}
\usepackage{url}
\usepackage{booktabs}
\usepackage{subfigure}
\usepackage{multirow}
\usepackage{fancybox}
\usepackage[dvipsnames,svgnames,x11names]{xcolor}
\usepackage{graphicx}
\usepackage{bbm}
\usepackage{cite}
\usepackage{epsfig, psfrag, amsmath, amssymb, amsfonts, latexsym, eucal, balance, indentfirst,bookmark, amsthm}

\newtheorem{thm}{Theorem}

\newtheorem{lem}{Lemma}

\newtheorem{rem}{Remark}

\makeatletter

\newcommand{\Rmnum}[1]{\expandafter\@slowromancap\romannumeral #1@}
\makeatother

\graphicspath{{figs/}}

\begin{document}

\title{Heterogeneous Cellular Networks with Spatio-Temporal Traffic: Delay Analysis and Scheduling}
\author{Yi Zhong, \emph{Member, IEEE}, Tony Q.S. Quek, \emph{Senior Member, IEEE} and Xiaohu Ge, \emph{Senior Member, IEEE}
\thanks{
Manuscript received Nov. 24, 2016.

Yi Zhong and Xiaohu Ge (corresponding author) are with
School of Electronic Information and Communications, Huazhong University
of Science and Technology, Wuhan, P. R. China. (e-mail: yzhong@hust.edu.cn,
xhge@mail.hust.edu.cn).
Tony Q.S. Quek is with Information Systems Technology and Design, Singapore University of Technology and Design, Singapore (email: tonyquek@sutd.edu.sg).

The authors would like to acknowledge the support from the NSFC Major International Joint Research Project (Grant No. 61210002), Hubei Provincial Science and Technology
Department under Grant 2016AHB006, the Fundamental Research Funds for the Central Universities under the grant 2015XJGH011. This research is partially supported by the EU FP7-PEOPLE IRSES, project acronym CROWN (grant no. 610524), China international Joint Research Center of Green Communications and Networking (No. 2015B01008).
This work was supported in part by the MOE ARF Tier 2 under Grant MOE2015-T2-2-104 and the Zhejiang Provincial Public Technology Research of China under Grant 2016C31063.
}
}
\maketitle
\begin{abstract}
Emergence of new types of services has led to various traffic and diverse delay requirements in fifth generation (5G) wireless networks.
Meeting diverse delay requirements
is one of the most critical goals for the design of 5G wireless networks.
Though the delay of point-to-point communications
has been well investigated, the delay of multi-point to multi-point
communications has not been thoroughly studied since it is a complicated function of all links in the network.
In this work, we propose a novel tractable approach to analyze the delay in the  heterogenous cellular networks with spatio-temporal random arrival of traffic. Specifically, we propose the notion of \emph{delay outage} and evaluated the effect of different scheduling policies on the delay performance.
Our numerical analysis reveals that offloading policy based on cell range expansion greatly reduces the macrocell traffic while bringing a small amount of growth for the picocell traffic. Our results also show that the delay performance of round-robin scheduling outperforms first in first out scheduling for heavy traffic, and it is reversed for light traffic.
In summary, this analytical framework provides an understanding and a rule-of-thumb for the practical deployment of 5G systems where delay requirement is increasingly becoming a key concern.
\end{abstract}
%

\section{Introduction}
\subsection{Motivation}
Evolution for high-speed data applications, such as high-quality wireless video streaming, social networking, and machine-to-machine communications, has led to the explosive growth of data traffic demand for 5G wireless networks.
It is envisioned that the average capacity requirement of 5G wireless networks might reach $25$ Gbps/km$^2$, which is $100$ times higher compared to the current system capacity.
Meanwhile, the delay requirements for different types of traffic become more diversified due to the emergence of new types of applications \cite{6824752}, such as latency-critical applications like command-and-control of drones, advanced manufacturing, and tactile Internet \cite{simsek20165g, fettweis2014tactile}.
As wireless networks evolve, the coupling between traffic and network services becomes increasingly strong. The tremendous traffic and its dynamic variations play increasingly crucial roles in affecting the delay performance of 5G wireless networks.  
Therefore, theoretical analysis of delay in wireless networks is imperative to guide us in our design and deployment.

\begin{figure}
\centering
\includegraphics[width=0.4\textwidth]{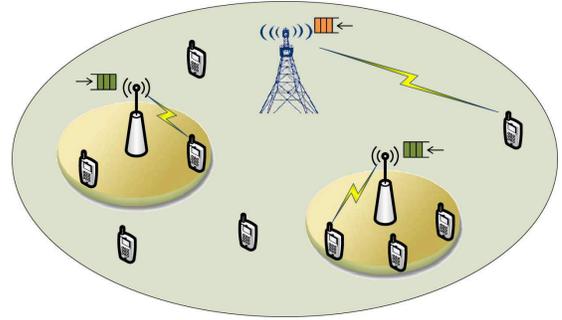}
\caption{Example for traffic in two-tier heterogeneous networks. The traffic is consist of the spatial distribution of users and the temporal arrival of packets.}
\label{fig:twocells}
\end{figure}

Though the delay of point-to-point communication has been well investigated using classical queueing theory, the delay of multi-point to multi-point communications, such as multi-party video conference or multi-player online gaming, 
has not been well studied. This is mainly due to the fact that delay is a complicated function of all links in the network and is affected by a variety of factors such as the traffic, the medium access protocol, the path loss, and so on.
In order to effectively analyze traffic and delay in heterogeneous cellular networks \cite{wildemeersch2013cognitive, cheung2012throughput, lopez2011enhanced, chen2015backhauling, lee2015hybrid, zhang2016delay}, a reasonable model to characterize both the spatial distribution and the temporal variation of traffic should first be established (see Fig. \ref{fig:twocells}).
The spatial distribution of traffic could be described by the locations of users. If the mobility of users is not considered, the temporal variation of traffic includes two parts, i.e., the random arrival process of packets for each user and the dynamic serving process of arrived packets.

\subsection{Related Works}
Most of the previous works only model one aspect of traffic, i.e., either only model the spatial distribution of users or only model the temporal arrival process of packets.
In \cite{zhong2013multi}, the distribution proposed in \cite{ferenc2007size} was used to approximate the probability density function (pdf) of the area for Voronoi cell formed by single-tier macrocell base stations (BSs).
With the obtained pdf, by modeling the locations of users as a homogeneous Poisson point process (PPP) \cite{haenggi2009stochastic, andrews2010tractable, baccelli2009stochastic2, haenggi2012stochastic}, the probability distribution of the number of users in each Voronoi cell was derived.
The works \cite{singh2013offloading, jo2012heterogeneous} extended such pdfs to the case of multi-tier heterogeneous networks.
On the other hand, the analysis of temporal variation of traffic were generally based on queueing theory and model the traffic as the function of random arrival of packets.
For example, in \cite{rao1988stability,tsybakov1979ergodicity,szpankowski1994stability,anantharam1991stability,luo1999stability}, a discrete-time slotted ALOHA system with multiple terminals was considered, and each terminal maintains a buffer of infinite capacity to store incoming packets.
The statuses (idle or busy) of terminals change over time due to the random arrival and serving of the traffic, and the serving rates also rely on the statuses of terminals since the idle terminals will not cause any interference.

\begin{figure}
\centering
\includegraphics[width=0.35\textwidth]{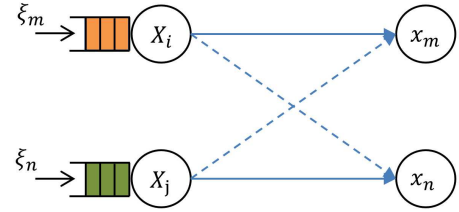}
\caption{Example of interacting queues problem of two cells in which the serving processes of the two links are coupled. For example, if one queue is empty, the corresponding transmitter do not cause interference to the other link; thus the serving rate of the other link will become large.}
\label{fig:InterQueues}
\end{figure}

Existing works that simultaneously model the spatio-temporal arrival of traffic, such as \cite{blaszczyszyn2015performance, abbas2015mobility, sapountzis2015analytical}, considered the traffic generated at random spatial regions, other than modeling the flow at each independent user.
The analysis in these works was based on the granularity of total traffic in each cell, i.e., the traffic generated in the coverage region of each BS is summed up as the traffic arrived at the corresponding BS.
For example, the work in \cite{blaszczyszyn2015performance} modeled the spatio-temporal arrival of users as a homogeneous spatio-temporal PPP, then the traffic arrived at each BS was characterized by the amount of data all users demand in the coverage of associated BS.
The main deficiencies of such a model are four aspects.
First, since the users are not definitely characterized in these models, many network operations, such as scheduling of users in each cell and offloading traffic between cells cannot be well captured. Meanwhile, the metric for each independent user, such as throughput and delay, cannot be well defined.
Second, due to the diversity of traffic, the quality of service (QoS) requirement and the arrival rate are different for different users, which cannot be appropriately modeled.
Third, the interference and the interaction between the queues at different BSs are either ignored or only analyzed by approximations. The interacting queues problem \cite{ephremides1987delay} is notoriously difficult to cope with since the statuses of queues and the serving rates are highly coupled (see Fig. \ref{fig:InterQueues}).
Lastly, these models are not consistent with those used in the industrial simulations \cite{access2010further}, in which the users are generated randomly and the packets arrive independently as a random process at each user.

Previous works using stochastic geometry to analyze the wireless networks focus on performance metrics such as success probability and average achievable rate by considering a snapshot of the network. However, these approaches are incapable to deal with long term metrics like delay, due to the difficulties such as the interacting queues problem, the spatial-temporal random arrival of traffic, and the static property of networks \cite{zhong2014managing,7486114}.
The works \cite{Wu2016Energy, Cheng2015Adaptive, Wu2016Delivering, Wu2016Energy2} analyzed the transmission of traffic with delay constraint in heterogeneous cellular networks without considering the coupling between traffic and network services.

\subsection{Contributions}
In this work, we model the spatio-temporal traffic by combining the tools from stochastic geometry and queueing theory.
Specifically, we model the spatial distribution of users by PPP and model the temporal arrival of packets at individual users by independent Bernoulli processes.
Then, the data flows for different users could be considered independently.
Meanwhile, since both the traffic and the delay requirements are diversified for different users, we model the packet arrival process and the delay requirements for each individual user. The main contributions of this paper can be summarized as follows:

\begin{itemize}
\item Several tractable approaches, for example, considering the dominant system and the modified system, are proposed to analyze and bound the statistical distribution of the signal-to-interference ratio (SIR) and the delay in heterogeneous cellular networks based on the combination of stochastic geometry and queuing theory.
\item Delay outage (DO) is proposed to characterize whether the delay requirement could be achieved. Furthermore, the effect of scheduling policies, i.e., the random scheduling, the first-input-first-output (FIFO) scheduling, and the round-robin scheduling, on the delay is evaluated, which to our knowledge has not been explored using stochastic geometry due to  analytical difficulty.
\item By numerical evaluations, statistics of traffic in heterogeneous cellular networks are investigated, and effect of spatial distribution and temporal variation of traffic on the delay is analyzed. Our proposed analytical framework provides a useful guideline for the system design and deployment of 5G wireless systems where the delay requirement becomes the key concern.
\end{itemize}

Our results reveal that the delay performance of round-robin scheduling outperforms FIFO scheduling for heavy traffic, and it is reversed for light traffic. Moreover, it shows that the offloading policy based on cell range expansion (CRE) greatly reduces the macrocell traffic while bringing a small amount of growth for the picocell traffic.

The remaining part of the paper is organized as follows.
Section \ref{sec:model} describes the spatial distribution model, the arrival process, and the scheduling policies.
Sections \ref{sec:traffic} and \ref{sec:delay} evaluate the statistic of traffic, success probability, and mean delay.
Section \ref{sec:numerical} numerically analyzes the effect of the spatial-temporal traffic on the statistical distribution of mean delay and the delay outage.
Finally, Section \ref{sec:conclusions} concludes the paper.

\section{System Model}
\label{sec:model}
We consider the downlink of a multi-tier heterogeneous cellular network with $K$ tiers of BSs. The locations of the $k$th tier ($k=1,...,K$) BSs are modeled as a homogeneous PPP $\Phi_k$ with intensity $\lambda_k$ BSs per km$^2$, and $\Phi_1, \Phi_2,..., \Phi_K$ are assumed to be independent.
Let $\Phi_b$ be the superposition of $\Phi_1, \Phi_2,..., \Phi_K$ as follows:
\begin{equation}
\Phi_b\triangleq\Phi_1\cup\Phi_2\cup...\cup\Phi_K,
\end{equation}
which denotes the locations of all tiers of BSs in the network.
The transmit power of the BSs of the $k$th tier is fixed as $P_k$.
We assume that the time is slotted into discrete time slots.

The traffic model in the network consists of two parts: the spatial distribution model of the users and the temporal arrival model of the packets at each user.
The mobility of the users is not modeled in the following, i.e. the temporal variation of the locations of the users are not considered.
However, a variety of mobility models, such as random walk model, random waypoint model, high mobility random walk model \cite{ge2016user}, could be applied to extend this work.
Therefore, our network model considered here is static since the locations of the BSs and the users remain static during all time after they are deployed \cite{net:Haenggi13tit, zhong2015stability, haenggi2015meta}.
We assume that the users are distributed as a homogeneous PPP $\Phi_u=\{x_i\}$ with intensity $\lambda_u$ users per m$^2$.
As for the temporal arrival of the downlink traffic at BSs, we assume that the packets arrival process for each user $x_i\in\Phi_u$ is an independent Bernoulli process of arrival rate $\xi_i$ (packets per time slot), i.e., $\xi_i$ is the probability of an arrival at user $x_i$ in any given time slot.
Without loss of generality, we assume that the size of each packet is fixed and it requires exactly one time slot to transmit a packet.
Each BS will maintain an independent queue of infinite size for each user within the coverage of associated BS to save the generated packets; thus the number of queues at a BS equals to the number of users served by the associated BS.
Due to diverse latency requirements in future wireless network, we assume that the mean delay requirement for user $x_i$ is $\beta_i$ (in number of time slots), i.e., the mean delay of user $x_i$ should be less than $\beta_i$ time slots.
With these notations, a user can be described by a triple $(x_i,\xi_i,\beta_i)$ with $x_i$ being the location, $\xi_i$ being the arrival rate of packets, and $\beta_i$ being the mean delay requirement.
In the following, we assume that $\{\xi_i\}$ and $\{\beta_i\}$ are independent identically distributed (i.i.d.) uniform random variables within $[\xi_{\min},\xi_{\max}]$ and $[\beta_{\min},\beta_{\max}]$, respectively.
\footnote{Note that our analysis can be extended to the case of general probability distributions for $\xi_i$ and $\beta_i$.}
Therefore, the traffic in the network can be modeled by a marked Poisson process $\widetilde{\Phi}_u$ defined as follows:
\begin{equation}
\widetilde{\Phi}_u \triangleq \{(x_i,\xi_i,\beta_i)\}.
\end{equation}

The propagation loss is modeled by two parts, the path loss and the fading.
A deterministic path loss function $l(r)=Ar^{-\alpha}$ is used to model the path loss, where $A>0$ is the reference path loss and $\alpha>2$ is the path loss exponent.
The fading is assumed to be Rayleigh block fading with unit mean, i.e., the power fading coefficients in different time slots are i.i.d. and are constant during one time slot.
We consider the interference-limited regime and ignore the thermal noise.
If the signal-to-interference ratio (SIR) at a user is larger than a certain threshold $\theta$, a packet will be successfully decoded.
Otherwise, if the acceptance of a packet fails, the packet will be added into the head of the queue and wait to be rescheduled again.
This model is based on the Shannon's theorem declaring that it is possible to communicate digital information nearly error-free up to certain rate determined by the SIR.
We summarized several crucial system parameters in Table \ref{tb:1}.

\begin{table}[tbp]
\centering
\caption{System parameters}
\label{tb:1}
\begin{tabular}{|c|c|c|c|}
\hline
Symbol & Description & Symbol & Description \\
\hline
  $K$ & \text{Number of tiers}   & $\Phi_k$ &  \text{Locations of $k$th tier BSs} \\
\hline
  $\lambda_u$ & \text{Density of users}   &
  $A$ & \text{Reference path loss} \\
  \hline
  $\alpha$ & \text{Path loss exponent}   & $\beta_i$ &  \text{Delay requirement for $x_i$} \\
  \hline
  $\theta$ &  \text{SIR threshold}   & $\Phi_u$ &  \text{Locations of the users} \\
  \hline
  $\xi_i$ &  \text{Arrival rate for $x_i$} & $P_k$ & \text{Power of $k$th tier BSs} \\
  \hline
\end{tabular}
\end{table}

The random arrival of traffic may greatly affect the performance of the network.
In order to improve the overall performance with random arrival of traffic, we consider several mechanisms to manage the traffic.
We first consider the CRE in which a bias factor for each tier of BSs is determined for offloading \cite{damnjanovic2011survey}.
By setting different bias factors for different tiers of BSs, the coverage region of the small cells can be extended to serve more users. Meanwhile, the number of users served by the macro BSs can then be reduced. Another mechanism is to improve the SIR at the users by mitigating the interference through randomly muting the transmission of the BSs in each time slot.
In practical wireless network, in the muted time slots, only control channels and cell-specific reference signals are transmitted with reduced power, and no user data is transmitted. During these time slots, the interference can be greatly reduced.
The last mechanism is to introduce the scheduling of users in each cell, i.e., the BS in each cell selects an appropriate user to serve.
Several scheduling mechanisms have been considered, such as
the random scheduling in which each user is selected to be served randomly with equal probability, the FIFO scheduling in which the packets are served in the manner of first come first serve, and the round-robin scheduling in which the users are scheduled one by one.

As for the user association strategy, we assume that the CRE is employed, where the association is based on the maximum average biased received power.
Let $B_k$ be the association bias for the $k$th tier of BSs. Thus, a user located at $x$ will associate the BS $X(x)$ denoted by
\begin{equation}
X(x)=\mathop{\arg\max}_{k\in\{1,2,...,K\}, y\in\Phi_k}B_kP_k|y-x|^{-\alpha}.
\end{equation}
By setting $B_k=1,\forall k$, the association strategy is exactly based on the maximum received power.
For the BSs from $k$th tier with small transmit power $P_k$, the association bias $B_k$ is set to be larger so that users can be offloaded from the overload macro cells to the light loaded small cells, and the overall interference in the network can also be reduced.
With these settings, the spatial tessellation of cells construes a weighted Voronoi tessellation.
Mathematically, the association region of a BS from the $k$th tier located at $X_k$ is given by
\begin{multline}
\mathcal{C}(X_k)=\bigg\{y\in \mathbb{R}^2: |y-X_k|\leq\bigg(\frac{B_kP_k}{B_iP_i}\bigg)^{1/\alpha}|y-z|, \\
\forall i\in\{1,2,...,K\},\forall z\in\Phi_i\bigg\}.
\end{multline}

As for the random muting, we assume that the blank time slots appear randomly at each BS with probability $1-p\ (0\leq p\leq 1)$ and independently among different BSs.
With these assumptions, in each time slot, each BS will suppress its transmission with probability $1-p$, thus postponing its own transmission but reducing its interference to the network.

Since there are multiple users served by each BS and only one user in a cell can be scheduled for transmission in each time slot, we consider three scheduling strategies as follows:

\subsubsection{Random Scheduling}
In random scheduling, each active BS will randomly choose one user within the coverage region of that BS to serve in each time slot.
Since there are multiple queues at each BS corresponding to multiple users served by that BS, each queue will be scheduled with the same probability in each time slot.
Due to the retransmission mechanism, if a packet fails for transmission in the scheduled time slot, the packet will be added into the head of the queue of the corresponding user and wait to be rescheduled again.

\subsubsection{FIFO Scheduling}
In FIFO scheduling, the packets arrived at each BS are served by the manner of FIFO, i.e., a packet arrived first will be scheduled first.
The result of FIFO scheduling only relies on the time when the packets arrive at a BS and is irrelevant to the particular user a packet belonging to.
Therefore, all queues at a BS can be considered as a large queue, a packet arrived first will be served first.

\subsubsection{Round-robin Scheduling}
In round-robin scheduling, the users served by each BS are scheduled one by one. For example, if there are $N$ users served by a BS, $N$ time slots will form a complete cycle in that cell, and the scheduling of each user will occupy one of the $N$ time slots.
Similarly, retransmission mechanism is introduced for the failed packets.
Note that each BS may suppress its transmission in a time slot with probability $1-p$ due to the random muting, the transmission of the scheduled user in that time slot will be suppressed.

Though several simple assumptions are used to make the analysis feasible, the analytical framework proposed in this work could be easily extended to model the particular scenarios and detailed technologies in wireless networks.

\section{Traffic Statistic}
\label{sec:traffic}
In this section, we evaluate the traffic statistic to characterize the spatio-temporal traffic, which is characterized by the spatial distribution, the temporal arrival rate, and the delay requirement for the users.
In order to characterize the traffic, we first introduce the following lemma.
\begin{lem}
\label{lem:1}
The probability that the typical user associates to the BSs of the $k$th tier is given by
\begin{equation}
\mathcal{P}_k = \frac{\lambda_k(P_kB_k)^{2/\alpha}}{\sum_{i=1}^K\lambda_i(P_iB_i)^{2/\alpha}}. \label{eqn:associateP}
\end{equation}
\end{lem}
\begin{proof}
The proof is similar to Lemma 1 in \cite{singh2013offloading}, and we present a simplified version here for completeness.
Let $Z_i$ be the distance between the typical user at the origin and the nearest BS in the $i$th tier of BSs.
The pdf of $Z_i$, given by
\begin{equation}
f_{Z_i}(z)=2\pi\lambda_iz\exp(-\pi\lambda_iz^2), \label{eqn:fZi}
\end{equation}
can be obtained by the void probability of the PPP in a ball.
The probability that the typical user associates to the BSs of the $k$th tier is
\begin{eqnarray}
\mathcal{P}_k &=& \mathbb{P}\left(\bigcap_{i=1,i\neq k}^{K}\{P_kB_kZ_k^{-\alpha}>P_iB_iZ_i^{-\alpha}\}\right)\nonumber\\
&\stackrel{(a)}{=}&\prod_{i=1,i\neq k}^{K}\mathbb{P}(P_kB_kZ_k^{-\alpha}>P_iB_iZ_i^{-\alpha}),
\end{eqnarray}
where $(a)$ follows from the independence of $\{\Phi_k\}$. Plugging in the pdf of $Z_i$, we obtain (\ref{eqn:associateP}).
\end{proof}

With the association probability given by Lemma \ref{lem:1}, we obtain the following lemma.

\begin{lem}
The pdf of the coverage area $S_k$ of a BS of the $k$th tier is approximated as
\begin{equation}
f_{S_k}(x)\simeq\frac{343}{15}\sqrt{\frac{3.5}{\pi}}\left(\frac{x\lambda_k}{\mathcal{P}_k}\right)^{2.5}\exp\left(-\frac{3.5x\lambda_k}{\mathcal{P}_k}\right)\frac{\lambda_k}{\mathcal{P}_k}. \label{eqn:pdfSk}
\end{equation}
\end{lem}
\begin{proof}
The pdf of the area $S$ of a Voronoi cell for a PPP with intensity $\lambda$ is \cite{zhong2013multi,ferenc2007size}
\begin{equation}
f_S(x)\simeq\frac{343}{15}\sqrt{\frac{3.5}{\pi}}(x\lambda)^{2.5}\exp\left(-3.5x\lambda\right)\lambda. \label{eqn:pdfSPPP}
\end{equation}
Due to the ergodicity of the PPP, since $\mathcal{P}_k$ is the probability that a user associates to the BSs of the $k$th tier, the average fraction of total area covered by the BSs of the $k$th tier is also $\mathcal{P}_k$.
Therefore, the average coverage area of a BS of the $k$th tier is ${\mathcal{P}_k}/{\lambda_k}$.
By replacing $\lambda$ with ${\lambda_k}/{\mathcal{P}_k}$ in (\ref{eqn:pdfSPPP}), we get the approximated result (\ref{eqn:pdfSk}).
\end{proof}

The following lemma gives the statistics for different kinds of users in each cell.
\begin{lem}
\label{lem:pgf}
Let $N_{k,\xi,\beta}$ be the number of users served by a BS of the $k$th tier with arrival rates less than $\xi\in[\xi_{\rm min},\xi_{\rm max}]$ and with delay requirements less than $\beta\in[\beta_{\rm min},\beta_{\rm max}]$.
The probability generating functions (PGFs) of $N_{k,\xi,\beta}$ are given by
\begin{equation}
G_{N_{k,\xi,\beta}}(z) \simeq \left(
1+C_0-C_0z\right)^{-\frac{7}{2}}. \label{eqn:pgf_xi}
\end{equation}
The probability mass functions (PMF) of $N_{k,\xi,\beta}$, denoted by $\mathbb{P}(N_{k,\xi,\beta}=n) = \frac{1}{n!}G_{N_{k,\xi,\beta}}^{(n)}(0)$, is given by
\begin{eqnarray}
\mathbb{P}(N_{k,\xi,\beta}=n) = \frac{(2n+5)!!}{15\cdot n!}\Big(\frac{C_0}{2}\Big)^n(C_0+1)^{-n-\frac{7}{2}}.
\end{eqnarray}
where $n!!$ is the double factorial and
\begin{eqnarray}
C_0=\frac{(\beta-\beta_{\rm min})(\xi-\xi_{\rm min})}{(\beta_{\rm max}-\beta_{\rm min})(\xi_{\rm max}-\xi_{\rm min})}\frac{\mathcal{P}_k}{\lambda_k}\frac{2}{7}\lambda_u. \label{eqn:C0}
\end{eqnarray}
\end{lem}
\begin{proof}
Since the spatial distribution of users construes a PPP with intensity $\lambda_u$,
given the area $S_k$ of a cell, the number of users within the cell with arrival rate less than $\xi$ and with delay requirement less than $\beta$ is a Poisson random variable with mean $\frac{(\beta-\beta_{\rm min})(\xi-\xi_{\rm min})\lambda_uS_k}{(\beta_{\rm max}-\beta_{\rm min})(\xi_{\rm max}-\xi_{\rm min})}$. Thus, the PGF of $N_{k,\xi,\beta}$, $G_{N_{k,\xi,\beta}}(z)=\mathbb{E}\left[z^{N_{k,\xi,\beta}}\right]$, is
\begin{eqnarray}
\!\!\!\!&\!\!\!\!&\!\!\!\!G_{N_{k,\xi,\beta}}(z) \nonumber\\ \!\!\!\!&\!\!=\!\!&\!\!\!\!\mathbb{E}_{S_k}\left[\exp\left(\frac{(\beta-\beta_{\rm min})(\xi-\xi_{\rm min})\lambda_uS_k(z-1)}{(\beta_{\rm max}-\beta_{\rm min})(\xi_{\rm max}-\xi_{\rm min})}\right)\right] \nonumber\\
\!\!\!\!&\!\!\simeq\!\!&\!\!\!\!\int_0^\infty\frac{343}{15}\sqrt{\frac{3.5}{\pi}}\left(\frac{\lambda_k}{\mathcal{P}_k}\right)^{3.5}x^{2.5}\nonumber\\
\!\!\!\!&\!\!\!\!&\!\!\exp\left(-\frac{3.5\lambda_k}{\mathcal{P}_k}x+\frac{(\beta-\beta_{\rm min})(\xi-\xi_{\rm min})\lambda_u(z-1)x}{(\beta_{\rm max}-\beta_{\rm min})(\xi_{\rm max}-\xi_{\rm min})}\right)\mathrm{d}x \nonumber\\
\!\!\!\!&\!\!=\!\!&\!\!\!\!\left(1-\frac{(\beta-\beta_{\rm min})(\xi-\xi_{\rm min})}{(\beta_{\rm max}-\beta_{\rm min})(\xi_{\rm max}-\xi_{\rm min})}\frac{\mathcal{P}_k}{\lambda_k}\frac{2}{7}\lambda_u(z-1)\right)^{-3.5}. \quad
\end{eqnarray}
The $n$th derivative of $G_{N_{k,\xi,\beta}}(z)$ is
\begin{eqnarray}
G_{N_{k,\xi,\beta}}^{(n)}(z)=\frac{C_0^n}{2^n}\frac{(2n+5)!!}{15}(-C_0z+C_0+1)^{-n-\frac{7}{2}},
\end{eqnarray}
where $C_0$ is given by (\ref{eqn:C0}) and $n!!$ is the double factorial.
Therefore, we get the PMF as follows
\begin{eqnarray}
\mathbb{P}(N_{k,\xi,\beta}=n) \!\!&\!\!=\!\!&\!\! \frac{1}{n!}G_{N_{k,\xi,\beta}}^{(n)}(0) \nonumber\\
\!\!&\!\!=\!\!&\!\!  \frac{(2n+5)!!}{15\cdot n!}\Big(\frac{C_0}{2}\Big)^n(C_0+1)^{-n-\frac{7}{2}}. \quad
\end{eqnarray}
\end{proof}

\begin{figure}
\centering
\includegraphics[width=0.45\textwidth]{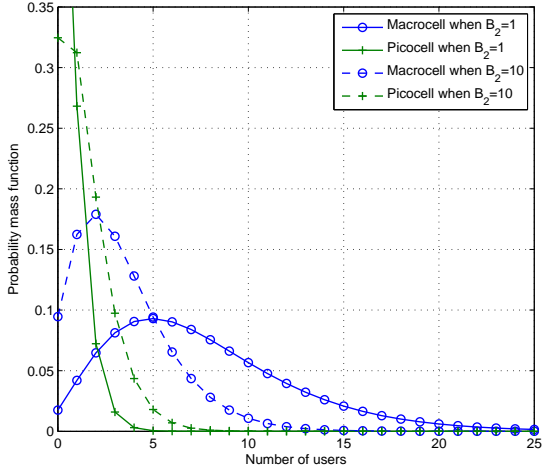}
\caption{PMFs of the number of users associating to different tiers of BSs in the two tiered heterogeneous networks formed by macrocells and picocells.
The transmit powers, the densities and the bias factors for macro BSs and micro BSs are set as $P_1=39$dBm and $P_2=24$dBm, $\lambda_1=0.00001$m$^{-2}$ and $\lambda_2 = 0.00005$m$^{-2}$, $B_1=1$ and $B_2=1$ or $10$, respectively. Other parameters are set as $\alpha=2.5$ and $\lambda_u=0.0001$m$^{-2}$.
}
\label{fig:PMF}
\end{figure}

Figure \ref{fig:PMF} plots the PMFs of the number of users associating to different tiers of BSs in the two tiered heterogeneous networks formed by macrocells and picocells. When the bias factors for macrocells and picocells are the same $(B_1=B_2=1)$, i.e., no bias exists, the macrocells will serve more users than the picocells because they provide larger coverage.
Figure \ref{fig:PMF} also indicates that in the case where some users are offloaded from the macrocells to the picocells $(B_1=1,B_2=10)$, the number of users served by each macrocell is greatly reduced while that served by each picocell does not increase much, implying that the offloading policy is potential for better utilization of the idle resources of the small cells.

\begin{rem}
From Lemma \ref{lem:pgf}, we get the mean number of users served by a BS of the $k$th tier with arrival rate less than $\xi$ and with delay requirement less than $\beta$ as
\begin{eqnarray}
\overline{N}_{k,\xi,\beta}\triangleq\mathbb{E}\left[{N_{k,\xi,\beta}}\right]= \frac{(\beta-\beta_{\rm min})(\xi-\xi_{\rm min})\mathcal{P}_k}{(\beta_{\rm max}-\beta_{\rm min})(\xi_{\rm max}-\xi_{\rm min})}\frac{\lambda_u}{\lambda_k}.
\end{eqnarray}
By setting $\xi=\xi_{\rm max}$ and $\beta=\beta_{\rm max}$, we get the total number of users served by a BS of the $k$th tier as $N_k=N_{k,\xi_{\rm max},\beta_{\rm max}}$,  whose PGF, PMF and mean value are
\begin{eqnarray}
&&G_{N_k}(z) \simeq \left(1-\frac{\mathcal{P}_k}{\lambda_k}\frac{2}{7}\lambda_u(z-1)\right)^{-3.5}, \label{eqn:pgfNk}\\
&&\mathbb{P}(N_k=n) = \frac{1}{n!}G_{N_{k}}^{(n)}(0), \label{eqn:pmfNk}\\
&&\overline{N}_{k}\triangleq\mathbb{E}\left[{N_{k}}\right]=\frac{\mathcal{P}_k}{\lambda_k}\lambda_u.
\end{eqnarray}
Letting $\xi_{k, \rm total}$ be the total arrival rate of the packets at a BS of the $k$th tier, we get
\begin{eqnarray}
\overline{\xi}_{k,\rm total}\triangleq\mathbb{E}\left[\xi_{k, \rm total}\right]= \frac{\xi_{\rm max}+\xi_{\rm min}}{2}\frac{\mathcal{P}_k}{\lambda_k}\lambda_u.
\end{eqnarray}
\end{rem}

\begin{figure}
\centering
\includegraphics[width=0.45\textwidth]{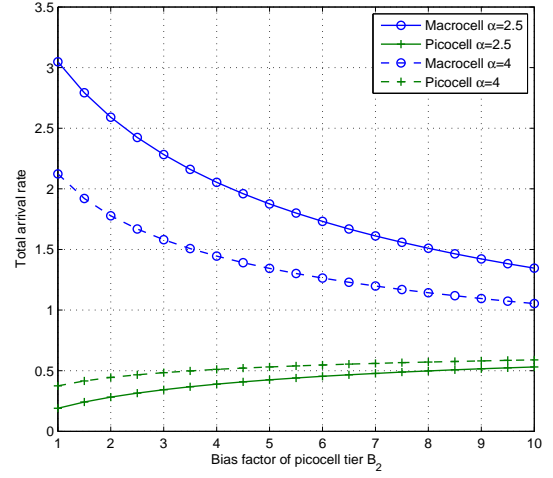}
\caption{Total arrival rates at different BSs in the two tiered heterogeneous networks formed by macrocells and picocells. The transmit powers and the densities  for macro BSs and micro BSs are set as $P_1=39$dBm and $P_2=24$dBm, $\lambda_1=0.00001$m$^{-2}$ and $\lambda_2 = 0.00005$m$^{-2}$, respectively. Other parameters are set as $B_1=1$, $\lambda_u=0.0001$m$^{-2}$, $\xi_{\rm min}=0.2$ and $\xi_{\rm max}=0.6$.}
\label{fig:Total_Arrival}
\end{figure}

Figure \ref{fig:Total_Arrival} plots the total arrival rate at different BSs in the two tiered heterogeneous networks formed by the macrocells and the picocells.
It is observed that when fixing the bias factor for macrocells and increasing the bias factor for picocells, the total arrival rate at each macrocell decreases dramatically while the total arrival rate at each picocell increases slowly, indicating that the traffic for macrocells is effectively offloaded with a little cost of increasing the traffic for each picocell.
Meanwhile, Figure \ref{fig:Total_Arrival} also reveals that the propagation environment may greatly affect the offloading of the traffic. For example,
the coverage region of the macrocells may shrink when increasing the path loss exponents $\alpha$, resulting in less traffic for the macrocells.

\section{Success Probability and Delay}
\label{sec:delay}
In the following, we analyze the statistics of success probability and delay in the heterogeneous networks.
The success probability is the probability that a scheduled packet can be successfully delivered.
The delay consists of two parts, one is the queueing delay, which measures the delay between the time when a packet arrives at the queue and the time when it starts to be served; the other is the service time, which is the time to transmit the given packet. We assume that the delay mentioned here is evaluated in terms of number of time slots.
Since the considered network is static, i.e., the locations of the BSs are deployed at first and keep unchanged during the following time, the
success probabilities for different links are different, resulting in different mean
delays for different queues.
Thus, a suitable metric to characterize the delay performance of the overall wireless network is the statistical cumulative distribution function (cdf) of the mean delay of all users in the network.
From the ergodicity of the PPP, the ensemble averages obtained by averaging over the point process equal the spatial averages obtained by averaging over an arbitrary realization of the PPP over a large region.
Therefore, in order to obtain the statistical cdf of the mean delay obtained from the spatial statistics of the mean delay samples of a large number of transmissions in the network, we could analytically evaluate the cdf of the mean delay at the typical transmission for different realizations of the PPP.

Without loss of generality, we consider a typical user located at the origin, denoted by $(x_0,\xi_0,\beta_0)$ with $x_0=o$, and a typical BS $y_0$ associated with the typical user.
For each realization of the spatial locations of BSs $\Phi_b$, we obtain a mean delay for the typical user.
By traversing all the realizations of $\Phi_b$, we obtain a statistical distribution of the mean delay (for example, the cdf) for the typical user at the origin.
Thus, the cdf of the mean delay for the typical user equals the statistical cdf of the mean delays of all users in the network.
To avoid confusion, we denote the mean delay of the typical user at the origin conditioning on the realization of $\Phi_b$ as the conditional mean delay, denoted by $\mathbf{D}_{\Phi_b}$, which it is a random variable uniquely defined by $\Phi_b$
\cite[Def. 1]{7448705}.
In the following discussions, we derive the cdf of the
conditional mean delay $\mathbf{D}_{\Phi_b}$, which is suitable to characterize the delay performance of the large  network since it also gives the statistical cdf of the mean delays of all users in the network.

The typical user will associate the BS from the $k$th tier with probability $\mathcal{P}_k$.
Let $Z_i$ be the distance between the typical user at the origin and the nearest BS in the $i$th tier of BSs.
Let $y_0\in\Phi_b$ be the serving BS; then $Z_k=|y_0|$ holds when a user associates to BS from the $k$th tier.
The following lemma gives the pdf of the link distance $|y_0|$ conditioned on the user associating to BS from the $k$th tier.

\begin{lem}
Conditioned on the typical user associating to BS $y_0$ from the $k$th tier, the pdf of the distance $Z_k=|y_0|$ between the typical user at the origin and its serving BS $y_0$ is
\begin{equation}
f_{Z_k\mid y_0\in\Phi_k}(r)=\frac{2\pi r\lambda_k}{\mathcal{P}_k}\exp\left(-\frac{\pi r^2\lambda_k}{\mathcal{P}_k}\right). \label{eqn:pdfnearest}
\end{equation}
\end{lem}
\begin{proof}
Conditioned on the typical user associating to BS from the $k$th tier, the cdf of the distance between the typical user and its serving BS $y_0$ is
\begin{eqnarray}
\!\!&\!\!\!\!&\!\!\mathbb{P}\left(|y_0|\leq r \mid y_0\in\Phi_k\right)\nonumber\\
\!\!&\!\!=\!\!&\!\!\frac{\mathbb{P}\left(|y_0|\leq r, y_0\in\Phi_k\right)}{\mathbb{P}\left(y_0\in\Phi_k\right)}\nonumber\\
\!\!&\!\!=\!\!&\!\!\frac{1}{\mathcal{P}_k}{\mathbb{P}\bigg(\{Z_k\leq r\}\bigcap\bigg\{\bigcap_{i=1,i\neq k}^{K}\{P_kB_kZ_k^{-\alpha}>P_iB_iZ_i^{-\alpha}\}\bigg\}\bigg)}\nonumber\\
\!\!&\!\!=\!\!&\!\!\frac{1}{\mathcal{P}_k}\int_0^r\prod_{i=1,i\neq k}^{K}\mathbb{P}(P_kB_kz^{-\alpha}>P_iB_iZ_i^{-\alpha})f_{Z_k}(z)\mathrm{d}z \nonumber\\
\!\!&\!\!=\!\!&\!\!1-\exp\left(-\pi r^2\sum_{i=1}^K\lambda_i\left(\frac{P_iB_i}{P_kB_k}\right)^{2/\alpha}\right),
\end{eqnarray}
where the last equation follows by plugging in the pdf of $Z_i$ given by (\ref{eqn:fZi}). By differentiating with respect to $r$, we get the result in the lemma.
\end{proof}

\begin{rem}
The form of the conditional pdf given by (\ref{eqn:pdfnearest}) is very similar to the contact distance distribution of the PPP given by (\ref{eqn:fZi}).
The only difference lies in that the intensity of the PPP $\lambda_i$ is replaced by $\frac{\lambda_k}{\mathcal{P}_k}$. Intuitively, it can be interpreted as that, since $\mathcal{P}_k$ is the average fraction of total area covered by BSs from the $k$th tier, the spatial distribution of the BSs from the $k$th tier could be considered as a PPP of intensity $\frac{\lambda_k}{\mathcal{P}_k}$.
\end{rem}

In the following, we first condition on the realization of $\Phi_b=\bigcup_{i=1}^K\Phi_i$ and derive the conditional mean delay $\mathbf{D}_{\Phi_b}$.
The statuses of BSs in the heterogeneous network switch between idle and busy according to the random muting, the statues of the queues, and the scheduling results at the BSs.
In each time slot, a BS will suppress its transmission and not cause interference if the BS is muted or the queue of the scheduled user of the BS is empty.
Let $\zeta_{y,t}\in\{0,1\}$ be the indicator showing whether the BS located at $y\in\Phi_b$ is idle $(\zeta_{y,t}=0)$ or busy $(\zeta_{y,t}=1)$ in the time slot $t$.
Then, if the typical user located at the origin associates to a BS $y_0$ from the $k$th tier, the SIR at the user in the time slot $t$ is
\begin{equation}
\mathrm{SIR}_t = \frac{P_kh_{y_0}|y_0|^{-\alpha}}{\sum_{i=1}^K\sum_{y\in\Phi_i\setminus B(o,z_{ik})}\zeta_{y,t}P_ih_y|y|^{-\alpha}},
\end{equation}
where $h_y$ is the power fading coefficient between the typical user and the BS located at $y$ with $h_y\sim\mathrm{Exp}(1)$, and $B(o,z_{ik})$ is a ball centered at the origin with radius $z_{ik}=\big(\frac{P_iB_i}{P_kB_k}\big)^{1/\alpha}|y_0|$.

The difficulty of the delay analysis lies in the fact that the SIR relies on the statuses of the queues at all BSs, and on the other hand the SIR at the users also affects the statuses of the queues.
Decoupling the SIR analysis and the statuses of the queues at all BSs is difficult; thus, we propose to bound the SIR as follows:
\begin{itemize}
  \item \textbf{Lower bound of SIR:}
  Consider a dominant system in which the typical user and its serving BS behave exactly the same as that in the original system; however, for other BSs in the dominant system, we assume that when the queues of the scheduled users at those BSs become empty, those BSs continue to transmit ``dummy'' packets, thus continuing to cause interference to other users no matter whether the scheduled queues are empty or not. The queue size at each BS in the dominant system will never be smaller than that in the original system, resulting in smaller SIR and larger delay. Therefore, the obtained SIR under these assumptions will be a lower bound for the SIR in the original system.

  \item \textbf{Upper bound of SIR:}
  Consider a modified system as follows: the typical user and its serving BS behave exactly the same as that in the original system; however, if an arrived packet at an interfering BS is not scheduled or failed for the transmission in current time slot, it will be dropped rather than being retransmitted. In this way, since the interference in the modified system is always smaller than that in the original system and the packets will not accumulate at the interfering BSs, the obtained SIR will be an upper bound for the SIR in the original system.

\end{itemize}

\subsection{Statistic of Success Probability}
By using the aforementioned bounding approaches, $\{\zeta_{y,t}\}$ become independent random variables and are not related to the time slot index $t$.
In these cases, the SIRs at the typical user in different time slots, denoted by $\{\mathrm{SIR}_t\}$, are independent random variables and also not related to the time slot index $t$. For simplicity and without ambiguity, we omit the subscript $t$ and use $\zeta_{y}$ and $\mathrm{SIR}$ when deriving these bounds.
Therefore, the success probability conditioned on $\Phi_b$ when the typical user is served by a BS $y_0$ from the $k$th tier with link distance $Z_k=|y_0|$ is given by
\begin{multline}
\mathbb{P}(\mathrm{SIR}>\theta\mid\Phi_b, y_0\in\Phi_k,Z_k=|y_0|)= \\ \mathbb{P}\bigg(\frac{P_kh_{y_0}|y_0|^{-\alpha}}{\sum_{i=1}^K\sum_{y\in\Phi_i\setminus B(o,z_{ik})}\zeta_{y}P_ih_y|y|^{-\alpha}} \\
>\theta\mid\Phi_b, y_0\in\Phi_k,Z_k=|y_0|\bigg).
\end{multline}
The probability of $\zeta_{y}=1$ relies on the bounding approaches and the scheduling strategies.
We first assume
\begin{equation}
\mathbb{P}(\zeta_{y}=1)=q,
\end{equation}
and analyze the success probability; then we determine the values of $q$ for different bounding approaches and scheduling strategies.

\begin{lem}
\label{lem:succprob}
If each interfering BS is active independently with probability $q$, i.e., $\mathbb{P}(\zeta_{y}=1)=q$, the cdf of success probability for the typical user conditioned on $\Phi_b$ when it associates to BS from $k$th tier, denoted by $\mathbb{P}(\mathrm{SIR}>\theta\mid\Phi_b, y_0\in\Phi_k)$, is
\begin{multline}
\mathbb{P}(\mathbb{P}(\mathrm{SIR}>\theta\mid\Phi_b, y_0\in\Phi_k)\leq u)
= \\
\frac{1}{2}+\frac{1}{\pi{\mathcal{P}_k}} \int_0^{\infty}\frac{1}{\omega}\mathrm{Im}\big\{u^{-j\omega}
\kappa^{-1}(\omega,k,q)\big\}
\mathrm{d}\omega.
\end{multline}
where $\kappa(\omega,k,q)$ is defined as
\begin{multline}
\kappa(\omega,k,q)=-2\sum_{i=1}^K\frac{\lambda_i}{\lambda_k}\Big(\frac{P_iB_i}{P_kB_k}\Big)^\delta \\
\times\int_0^1\Big(1-\Big(\frac{q}{1+\theta y^\alpha\frac{B_k}{B_i}}+1-q\Big)^{j\omega}\Big)y^{-3}\mathrm{d}y -\frac{ 1}{\mathcal{P}_k} \\
=\delta\sum_{i=1}^K\frac{\lambda_i}{\lambda_k} \sum_{n=1}^\infty \binom{j\omega}{n}\Big(\frac{B_i}{B_k}\Big)^{\delta-n} \Big(\frac{P_i}{P_k}\Big)^\delta  \\ \frac{(-q\theta)^n}{n-\delta}{_2F_1\Big(n,n-\delta;n-\delta+1;-\theta\frac{B_k}{B_i}\Big)}-\frac{ 1}{\mathcal{P}_k}.
\end{multline}
\end{lem}
\begin{proof}
See Appendix \ref{appendix:a}.
\end{proof}

\begin{figure}
\centering
\includegraphics[width=0.45\textwidth]{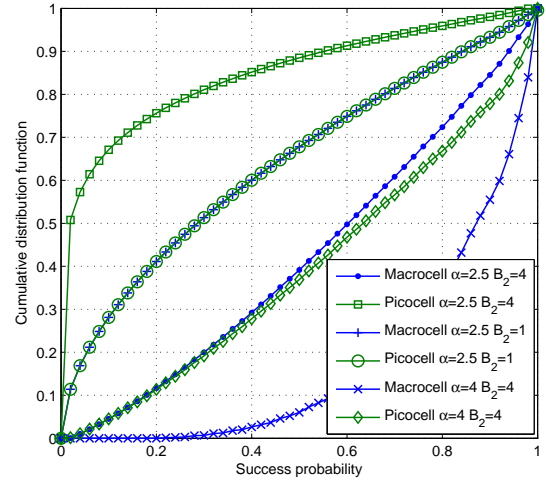}
\caption{Cdf of success probabilities for different tiers of BSs in the two tiered heterogeneous networks formed by macrocells and picocells. The transmit powers and the densities for macro BSs and micro BSs are set as $P_1=39$dBm and $P_2=24$dBm, $\lambda_1=0.00001$m$^{-2}$ and $\lambda_2 = 0.00005$m$^{-2}$, respectively. The bias factor for macrocell is $B_1=1$.}
\label{fig:SuccessProb}
\end{figure}

Figure \ref{fig:SuccessProb} plots the cdf of success probabilities for different tiers of BSs in the two tiered heterogeneous networks formed by macrocells and picocells. If there is no bias $(B_1=B_2=1)$, the cdfs of the success probabilities for macrocell users and picocell users coincide, implying that the statistical success probability of users associating to different tiers of BSs is the same for the maximum average received power association. The cdfs given by Figure \ref{fig:SuccessProb} shows that when increasing the bias factor for picocells $(B_1=1, B_2=4)$, the proportion of picocell users with small success probability increases since the offloaded users are distributed at the edge of the picocells and may experience poor SIR performance.
For the case of large path loss exponent, the electromagnetic signal attenuates fast, thus suppressing the inter-cell interference and increasing the success probability for both macrocell and picocell users. This observation indicates that offloading is more effective when the signal attenuates fast.

Define the success probability of the typical user conditioned on $\Phi_b$ as
\begin{equation}
\mathbf{P}_{\Phi_b}\triangleq\mathbb{P}(\mathrm{SIR}>\theta\mid\Phi_b).
\end{equation}
Then, $\mathbf{P}_{\Phi_b}$ is a random variable uniquely determined by the realization of $\Phi_b$.
By traversing all the realizations of $\Phi_b$, we get the cdf of $\mathbf{P}_{\Phi_b}$.
Due to the ergodicity of the PPP, the cdf of $\mathbf{P}_{\Phi_b}$ also gives the statistical cdf of the success probabilities of all users within a large region.
Therefore, we get the following theorem.

\begin{lem}
\label{thm:succprob}
If each interfering BS is active independently with probability $q$, i.e., $\mathbb{P}(\zeta_{y}=1)=q$, the statistical cdf of success probabilities of all users in the network equals the cdf of the success probability of the typical user conditioned on $\Phi_b$, which is
\begin{multline}
\mathbb{P}\left(\mathbf{P}_{\Phi_b}\leq u\right)
= F(u,q) \\
\triangleq\frac{1}{2}+\frac{1}{\pi}\sum_{k=1}^K \int_0^{\infty}\frac{1}{\omega}\mathrm{Im}\big\{u^{-j\omega}\kappa^{-1}(\omega,k,q)
\big\}\mathrm{d}\omega.
\end{multline}
\end{lem}
\begin{proof}
Combining Lemma \ref{lem:succprob} and using the total probability formula
\begin{equation}
\mathbb{P}\left(\mathbf{P}_{\Phi_b}\leq u\right)
=\sum_{k=1}^K \mathcal{P}_k\mathbb{P}(\mathbb{P}(\mathrm{SIR}>\theta\mid\Phi_b, y_0\in\Phi_k)\leq u),
\end{equation}
we obtain the theorem.
\end{proof}

\begin{rem}
Particularly, in single tier cellular network, i.e., $K=1$, the statistical cdf of success probabilities of all users in the network is
\begin{multline}
\mathbb{P}\left(\mathbf{P}_{\Phi_b}\leq u\right)
=\frac{1}{2}+ \frac{1}{\pi }\int_0^{\infty}\frac{1}{\omega}\mathrm{Im}\bigg\{ \\
\frac{u^{-j\omega}}
{\delta \sum_{n=1}^\infty \binom{j\omega}{n}   \frac{(-q\theta)^n}{n-\delta}{_2F_1(n,n-\delta;n-\delta+1;-\theta)}-1}\bigg\}
\mathrm{d}\omega.
\end{multline}

The cdf of the success probability given by Lemma \ref{thm:succprob}, $\mathbb{P}\left(\mathbf{P}_{\Phi_b}\leq u\right)$, equals the proportion of number of users whose success probabilities are less than $u$ in the network.
Lemma \ref{thm:succprob} reveals that, in the $K$-tier heterogeneous network, the cdf of the success probability $\mathbf{P}_{\Phi_b}$ only relies on the ratios of the densities of BSs, the bias factors and the transmit powers among different tiers.
For the single tier network ($K=1$), the cdf of the success probability is not related to the densities of BSs, the bias factors or the transmit powers of BSs.
These observations indicate that the statistical distribution of the success probabilities of all users in the single-tiered interference-limited heterogeneous network with much randomness in the deployment is only slightly correlated to the exact value of a number of system parameters, such as the densities of BSs or the transmit powers of BSs.
This result is coincident with the conclusion for the coverage probability in the networks modeled by the PPPs \cite{ge20165g}.
\end{rem}
%

\subsection{Statistic of Delay}
In the following, we discuss the statistics of the delay for different scheduling policies.

\subsubsection{Random Scheduling}
In the random scheduling, each active BS will randomly choose one user from all users within the coverage region of that BS to serve in each time slot.
Let $N$ be the number of users served by the BS.
Considering the random muting mechanism that each BS is idle with probability $1-p$ in each time slot, the average service rate (number of packets transmitted successfully per time slot) of the typical user conditioned on $\Phi_b$ is
\begin{equation}
\mu = \frac{p}{N}\mathbb{P}(\mathrm{SIR}>\theta\mid\Phi_b)=\frac{p}{N}\mathbf{P}_{\Phi_b}. \label{eqn:randommu}
\end{equation}

The probability for a packet at the typical user being scheduled and successfully transmitted in a time slot is $\mu$ given by (\ref{eqn:randommu}).
Since the transmissions of the typical user in different time slots are independent in the random scheduling, the probability for successfully transmitting a packet in any time slot is also $\mu$.
Therefore, the queueing system at the typical transmitter is equivalent to a Geo/G/1 queue, or a discrete-time single server retrial queue \cite{atencia2004discrete}, \cite{falin1990survey}. In the equivalent Geo/G/1 queue, the time is slotted and the packets arrive according to a Bernoulli process with intensity $\xi_0$ packets per time slot. The arrival process is also called geometric arrival process since the probability that a packet arrives in a time slot is $\xi_0$, and the number of time slots between two adjacent arrivals is a geometric random variable. The success probability is $\mu$ and the service times of packets are i.i.d. with geometric distribution. From \cite{atencia2004discrete}, we get the conditional mean delay $\mathbf{D}_{\Phi_b}$ in random scheduling as
\begin{eqnarray}
\mathbf{D}_{\Phi_b}= \left\{ \begin{array}{ll}
\frac{1-\xi_0}{\mu-\xi_0} & \textrm{if $\mu> \xi_0$}\\
\infty & \textrm{if $\mu\leq \xi_0$}.
\end{array} \right. \label{eqn:randomdelay}
\end{eqnarray}
Therefore, we obtain the following theorem.
\begin{lem}
\label{thm:rsdelay}
In random scheduling, if each interfering BS is active independently with probability $q$, i.e., $\mathbb{P}(\zeta_{y}=1)=q$, the statistical cdf of the mean delay of all users in the network equals the cdf of mean delay of the typical user conditioned on $\Phi_b$, which is
\begin{multline}
\mathbb{P}\left(\mathbf{D}_{\Phi_b}\leq T\right)
= H_1(T,q)
\triangleq \frac{1}{2}-\frac{1}{\pi} \sum_{k=1}^K  \int_0^{\infty}\frac{1}{\omega}\mathrm{Im}\Big\{\\
\frac{{p}^{j\omega}}{\kappa(\omega,k,q)}
\mathbb{E}_{N_k,\xi_0}\Big[{N_k^{-j\omega}\Big( \frac{1}{T}+\frac{T-1}{T} \xi_0\Big)}^{-j\omega}\Big]
\Big\}\mathrm{d}\omega.
\end{multline}
\end{lem}
\begin{proof}
See Appendix \ref{appendix:b}.
\end{proof}

Particularly, by the previous mentioned bounding approaches, we obtain the following theorem which gives the bounds for the cdf of success probability and mean delay in random scheduling.
\begin{thm}
\label{cor:rsdelay}
In random scheduling, the statistical cdf of success probability and mean delay of all users in the network are bounded by
\begin{eqnarray}
F(u,p) \leq \mathbb{P}\left(\mathbf{P}_{\Phi_b}\leq u\right) \leq F(u,\frac{\xi_{\rm max}+\xi_{\rm min}}{2}p), \\
H_1(T, p) \leq \mathbb{P}\left(\mathbf{D}_{\Phi_b}\leq T\right) \leq H_1(T, \frac{\xi_{\rm max}+\xi_{\rm min}}{2}p).
\end{eqnarray}
\end{thm}
\begin{proof}
By considering the aforementioned dominant system ($q=p$), we obtain a lower bound of SIR, which results in lower bounds for the success probability and the cdf of the mean delay. By considering the modified system ($q=\frac{\xi_{\rm max}+\xi_{\rm min}}{2}p$), we obtain an upper bound of SIR, which results in upper bounds for the success probability and the cdf of the mean delay. 
\end{proof}
%

\subsubsection{FIFO Scheduling}
In the FIFO scheduling, each active BS will schedule the earliest arriving packet at that BS in each time slot, and all queues at a BS could be considered as a single large queue.
Notice that the delay of the typical user equals the delay of the packets arriving at the large queue.
The average service rate of the large queue conditioned on $\Phi_b$ is
\begin{equation}
\mu = p\mathbb{P}(\mathrm{SIR}>\theta\mid\Phi_b)=p\mathbf{P}_{\Phi_b}. \label{eqn:fifomu}
\end{equation}

Generally, the probability that more than two packets arrive simultaneously at the large queue is very small.
Thus, the arrival process at the large queue can be considered as Bernoulli arrival with arrival rate $\sum_{i=0}^{N-1}\xi_i$, where $N$ is the number of users (including the typical user at $x_0$) and $\xi_i$ is the arrival rate of the $i$th user.
The queueing process at the BS is equivalent to a Geo/G/1 queueing system.
Therefore, the conditional mean delay $\mathbf{D}_{\Phi_b}$ in FIFO scheduling is
\begin{eqnarray}
\mathbf{D}_{\Phi_b}= \left\{ \begin{array}{ll}
\frac{1-\sum_{i=0}^{N-1}\xi_i}{\mu-\sum_{i=0}^{N-1}\xi_i}
& \textrm{if $\mu> \sum_{i=0}^{N-1}\xi_i$}\\
\infty & \textrm{if $\mu\leq \sum_{i=0}^{N-1}\xi_i$}.
\end{array} \right. \label{eqn:fifodelay}
\end{eqnarray}

Therefore, we obtain the following lemma and theorem.
\begin{lem}
\label{thm:fifodelay}
In FIFO scheduling, if each interfering BS is active independently with probability $q$, i.e., $\mathbb{P}(\zeta_{y}=1)=q$, the statistical cdf of the mean delay of all users in the network equals the cdf of the mean delay of the typical user conditioned on $\Phi_b$, which is
\begin{multline}
\mathbb{P}\left(\mathbf{D}_{\Phi_b}\leq T\right)
= H_2(T,q)
\triangleq\frac{1}{2}-\frac{1}{\pi} \sum_{k=1}^K  \int_0^{\infty}\frac{1}{\omega}\mathrm{Im}\Big\{\\
\frac{{p}^{j\omega}}{\kappa(\omega,k,q)}\mathbb{E}_{N_k,\{\xi_i\}}\Big[\Big(\frac{1}{T}+\frac{T-1}{T}\sum_{i=0}^{N_k-1}\xi_i\Big)^{-j\omega}\Big]\Big\}
\mathrm{d}\omega.
\end{multline}
\end{lem}
\begin{proof}
The proof is similar to that of Lemma \ref{thm:rsdelay}.
\end{proof}

\begin{thm}
\label{cor:fifodelay}
In FIFO scheduling, the statistical cdf of success probability and mean delay of all users in the network are bounded by
\begin{multline}
F(u,p) \leq \mathbb{P}\left(\mathbf{P}_{\Phi_b}\leq u\right) \\
\leq F\Big(u,  \min\Big\{1,\frac{p}{2}(\xi_{\rm max}+\xi_{\rm min}) \min_{1\leq k\leq K}\overline{N}_{k}\Big\} \Big),
\end{multline}
\begin{multline}
H_2(T, p) \leq \mathbb{P}\left(\mathbf{D}_{\Phi_b}\leq T\right) \leq \\
H_2\Big(T, \min\Big\{1,\frac{p}{2}(\xi_{\rm max}+\xi_{\rm min}) \min_{1\leq k\leq K}\overline{N}_{k}\Big\}\Big).
\end{multline}
\end{thm}
\begin{proof}
The proof is similar to that of Theorem \ref{cor:rsdelay}. The upper bound is obtained by choosing the minimum active probability of all tiers of BSs.
\end{proof}


\subsubsection{Round-robin Scheduling}
In the round-robin scheduling, if there are $N$ users within the typical cell, a complete cycle will occupy $N$ time slots. Therefore, the duration between two scheduled time slots for the typical user is $N$.
If we only consider the time slots that the typical user is scheduled, the serving of the typical user could be considered as a new queueing system with the serving rate given by
\begin{equation}
\mu = p\mathbb{P}(\mathrm{SIR}>\theta\mid\Phi_b)=p\mathbf{P}_{\Phi_b}. \label{eqn:rrmu}
\end{equation}

The new queueing system is also a Geo/G/1 queueing system.
Let $D_{\rm RR}$ be the mean delay of the new queueing system, which is the mean number of scheduled time slots required for the typical user to successfully transmit a packet.
Then, we have
\begin{eqnarray}
D_{\rm RR}= \left\{ \begin{array}{ll}
\frac{1-\xi_0}{\mu-\xi_0} & \textrm{if $\mu> \xi_0$} \\
\infty & \textrm{if $\mu\leq \xi_0$}.
\end{array} \right.
\end{eqnarray}

For the typical user, note that $N/2$ time slots are required on average if a packet is successfully delivered in the first scheduled time slot; otherwise, the additional time slots required should be the additional number of scheduled time slots multiplied by $N$. Therefore, the conditional mean delay $\mathbf{D}_{\Phi_b}$ in round-robin scheduling is
\begin{eqnarray}
\mathbf{D}_{\Phi_b} &=& N(D_{\rm RR}-1)+\frac{N}{2}\nonumber\\
&=& \left\{ \begin{array}{ll}
\frac{1-\xi_0}{\mu-\xi_0}N-\frac{N}{2} & \textrm{if $\mu> \xi_0$} \\
\infty & \textrm{if $\mu\leq \xi_0$}.
\end{array} \right. \label{eqn:rrdelay}
\end{eqnarray}

The following lemma and theorem give the bounds for the cdf of the conditional mean delay $\mathbf{D}_{\Phi_b}$ in round-robin scheduling.
\begin{lem}
\label{thm:fifodelay}
In round-robin scheduling, if each interfering BS is active independently with probability $q$, i.e., $\mathbb{P}(\zeta_{y}=1)=q$, the statistical cdf of the mean delay of all users in the network equals the cdf of the mean delay of the typical user conditioned on $\Phi_b$, which is
\begin{multline}
\mathbb{P}\left(\mathbf{D}_{\Phi_b}\leq T\right)
= H_3(T,q)
\triangleq\frac{1}{2}-\frac{1}{\pi} \sum_{k=1}^K  \int_0^{\infty}\frac{1}{\omega}\mathrm{Im}\Big\{\\
\frac{{p}^{j\omega}}{\kappa(\omega,k,q)}\mathbb{E}_{N_k,\xi_0}\Big[
\Big(\xi_0+\frac{1-\xi_0}{1/2+T/N_k}\Big)^{-j\omega}\Big]\Big\}
\mathrm{d}\omega.
\end{multline}
\end{lem}
\begin{proof}
The proof is similar to that of Lemma \ref{thm:rsdelay}.
\end{proof}

\begin{thm}
\label{cor:rrdelay}
In round-robin scheduling, the statistical cdf of success probability and mean delay of all users in the network are bounded by
\begin{eqnarray}
F(u,p) \leq \mathbb{P}\left(\mathbf{P}_{\Phi_b}\leq u\right) \leq F(u,\frac{\xi_{\rm max}+\xi_{\rm min}}{2}p), \nonumber\\
H_3(T, p) \leq \mathbb{P}\left(\mathbf{D}_{\Phi_b}\leq T\right) \leq H_3(T, \frac{\xi_{\rm max}+\xi_{\rm min}}{2}p).
\end{eqnarray}
\end{thm}
\begin{proof}
The proof is similar to that of Theorem \ref{cor:rsdelay}.
\end{proof}
%


Other user scheduling mechanisms, such as FDMA, could also be analyzed similarly. In the case of FDMA, all users served by a same BS will be served simultaneously using orthogonal sub-channels. Since not all sub-channels are used at a BS, the set of BSs that cause interference to the typical link will be only a subset of all BSs, i.e., the interfering BSs construe a thinning version of the original BSs. Therefore, the approaches used in our work can be also applied in the FDMA case, with the difference being that the interference is thinned, and the user scheduling is simplified.

\subsection{Delay Outage}
In the design of practical wireless network, an intuitive and meaningful metric is the proportion of users whose delay requirements cannot be achieved.
The delay requirement of a user $x_i$ cannot be achieved if the practical mean delay of the user is larger than the mean delay requirement $\beta_i$ of the user.
Therefore, we propose the notion of \emph{delay outage} which means that the delay requirement of a user cannot be achieved.
The user $x_i$ is delay outage if and only if its mean delay is larger than $\beta_i$.
We further evaluate the delay outage probability, which is the proportion of users that are delay outage in the wireless network, as
\begin{eqnarray}
\eta_\mathrm{DO}&\triangleq&\mathbb{E}_{\beta_i}\left(\mathbb{P}\left(\mathbf{D}_{\Phi_b}\geq \beta_i\right)\right)\nonumber\\
&=&\frac{1}{\beta_{\max}-\beta_{\min}}\int_{\beta_{\min}}^{\beta_{\max}}\mathbb{P}\left(\mathbf{D}_{\Phi_b}\geq \beta_i\right)\mathrm{d}\beta_i. \label{eqn:do}
\end{eqnarray}
The previous results could be plugged into (\ref{eqn:do}) to bound the delay outage probability $\eta_\mathrm{DO}$.


\section{Numerical Evaluation}
\label{sec:numerical}
To gain insight, we discuss the statistical cdf of mean delay and the delay outage numerically.

\begin{figure}
\centering
\includegraphics[width=0.45\textwidth]{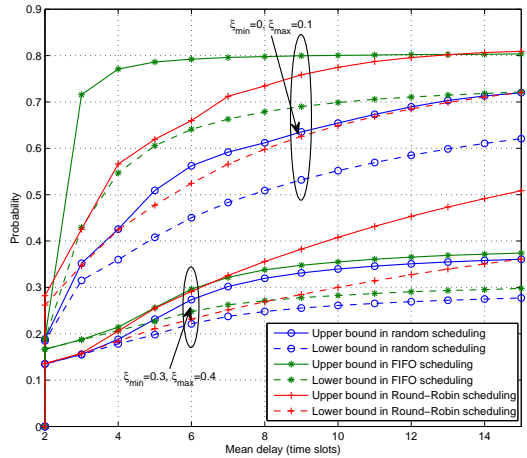}
\caption{Upper and lower bounds for the cdfs of the mean delay of all queues in the two tiered heterogeneous networks formed by macrocells and picocells. The parameters are set as $P_1=39$dBm, $P_2=24$dBm, $\lambda_1=0.00001$m$^{-2}$, $\lambda_2 = 0.00005$m$^{-2}$, $B_1=B_2=1$, $p=0.5$ and $\lambda_u=0.00005$m$^{-2}$.}
\label{fig:Delat_cdf}
\end{figure}

Figure \ref{fig:Delat_cdf} plots the upper and lower bounds for the statistical cdfs of the mean delay of all queues in the heterogeneous networks formed by macrocells and picocells.
It is observed that the bounds are tight for large arrival rates of the packets, which can be interpreted as that our proposed simplified system and dominant system when deriving the bounds tend to be the same system for large arrival rates.
Comparing the curves for different arrival rates, we observe that the delay performance of random scheduling is the worst while the delay performance of round-robin scheduling outperforms FIFO scheduling when the traffic is heavy, and it is reversed for light traffic.
This is because when the traffic is light, applying round-robin scheduling may waste some time slots to schedule the empty queues, and when the traffic is heavy,
in the FIFO scheduling, the scheduling may be blocked at the users with poor link quality, resulting in large delay for the packets.
Figure \ref{fig:Delat_cdf} shows that the cdf of the mean delay will not increase to one when increasing the value of the mean delay, due to the fact that some users may be affected by strong interference, and their queues increase to infinite resulting in infinite mean delay. We also observe that certain percentage of users in the network experience very low mean delay, i.e., the mean delay for those users approaches to two ($1/p$) time slots, indicating that there are always certain percentage of users that experience a small interference and a low mean delay.

\begin{figure}
\centering
\includegraphics[width=0.44\textwidth]{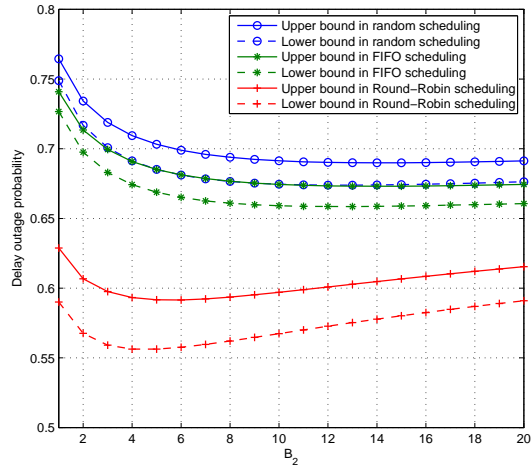}
\caption{Upper and lower bounds for the delay outage probability as functions of bias factor $B_2$ in the two tiered heterogeneous networks formed by macrocells and picocells. The parameters are set as $P_1=39$dBm, $P_2=24$dBm, $\lambda_1=0.00001$m$^{-2}$, $\lambda_2 = 0.00005$m$^{-2}$, $\xi_{\rm min}=0.7$, $\xi_{\rm max}=0.8$,  $\beta_{\rm min}=18$, $\beta_{\rm max}=20$, $B_1=1$, $p=1$ and $\lambda_u=0.00005$m$^{-2}$.}
\label{fig:vary_B2}
\end{figure}

Figure \ref{fig:vary_B2} plots the upper and lower bounds for the delay outage probability as functions of the bias factor for the picocells $B_2$ in the two tiered heterogeneous networks formed by macrocells and picocells for heavy traffic.
For all scheduling policies, when beginning to increase $B_2$ from $1$, the delay outage probability in the network decreases, indicating that offloading helps to reduce the traffic in the macrocells and effectively utilize the resources in the picocells. Meanwhile, it is observed that as $B_2$ continues to increase, the delay outage probability of random scheduling and FIFO scheduling tends to be constant while that of round-robin scheduling first decreases then increases. It can be interpreted as that in the case of random scheduling and FIFO scheduling, due to the heavy traffic, the scheduling for the macrocells and several picocells experiencing poor channel quality may be blocked and resulting in large delay, and only those users served by the picocells with good channel quality satisfy the delay requirement. However, for the round-robin scheduling, all users are ensured to be served in each scheduling period. If most of the users are offloaded from the macrocells to the picocells, the resources at the macrocells cannot be fully utilized, and the delay outage probability will start to increase for large $B_2$.

\begin{figure}
\centering
\includegraphics[width=0.45\textwidth]{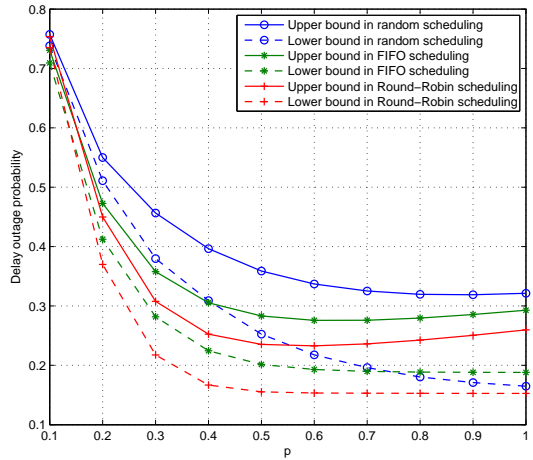}
\caption{Upper and lower bounds for the delay outage as functions of active probability $p$ in the two tiered heterogeneous networks formed by macrocells and picocells. The parameters are set as $P_1=39$dBm, $P_2=24$dBm, $\lambda_1=0.00001$m$^{-2}$, $\lambda_2 = 0.00005$m$^{-2}$, $B_1=B_2=1$, $\xi_{\rm min}=0$, $\xi_{\rm max}=0.1$, $\beta_{\rm min}=18$, $\beta_{\rm max}=20$, and $\lambda_u=0.00005$m$^{-2}$.}
\label{fig:vary_p}
\end{figure}

\begin{figure}
\centering
\includegraphics[width=0.45\textwidth]{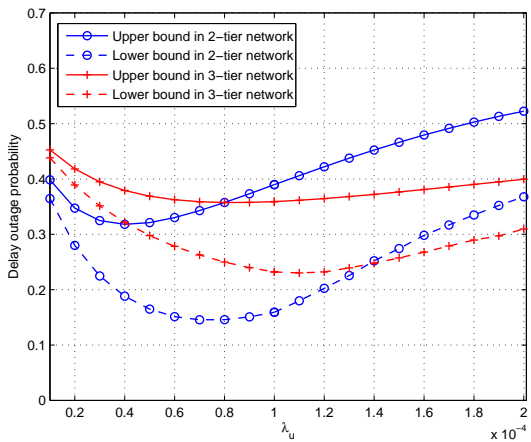}
\caption{Upper and lower bounds for the delay outage as functions of user density $\lambda_u$ for different number of tiers in heterogeneous networks. The parameters are set as $P_1=39$dBm, $P_2=24$dBm, $P_3=20$dBm, $\lambda_1=0.00001$m$^{-2}$, $\lambda_2 = 0.00005$m$^{-2}$, $\lambda_3 = 0.0001$m$^{-2}$, $B_1=B_2=B_3=1$, $\xi_{\rm min}=0$, $\xi_{\rm max}=0.1$, $\beta_{\rm min}=18$, and $\beta_{\rm max}=20$.}
\label{fig:difftier}
\end{figure}

Figure \ref{fig:vary_p} plots the upper and lower bounds for the delay outage probability as functions of the active probability $p$ in the two tiered heterogeneous networks formed by macrocells and picocells for light traffic ($\xi_{\rm min}=0$, $\xi_{\rm max}=0.1$). Generally, increasing $p$ will decrease the delay outage probability since the opportunity to schedule a packet is increased for large $p$; meanwhile, the interference will not increase much due to the light traffic.
It is observed that the bounds are tight for small $p$ and loose for large $p$.
This is because when deriving the upper bound, we introduced the dominant system where the interfering transmitters are assumed to be backlogged.
The busy probability of the BSs in the dominant system for small $p$ is much closer to the busy probability in the original system than that for large $p$.

Figure \ref{fig:difftier} plots the upper and lower bounds for the delay outage probability as functions of user density $\lambda_u$ for different number of tiers in heterogeneous cellular networks.
The figure reveals that the delay outage probability first decreases then increases as the increasing of user density.
This counterintuitive result can be interpreted as follows. When the user density $\lambda_u$ is much smaller than the density of BSs, many cells are empty and serve no user. Since we only consider the cells serving at least one user, slightly increasing $\lambda_u$ will increase the proportion between the number of cells serving one user and that of cells serving more than one users.
Noting that the cells with only one user experience much smaller delay outage probability than that with more than one users, the overall delay outage probability will decrease when slightly increasing the user density $\lambda_u$ in the case where $\lambda_u$ is much smaller than the density of BSs.
Figure \ref{fig:difftier} also reveals that deploying more tiers of BSs helps to decrease the delay outage probability only when the user density is large.

\section{Conclusion}
\label{sec:conclusions}
In this paper, a tractable approach is proposed to analyze the delay in the heterogeneous cellular networks with spatio-temporal random arrival of traffic.
Meanwhile, the statistic of spatio-temporal traffic with offloading policy is evaluated.
The notion of delay outage is proposed to evaluate the effect of different scheduling policies, such as random scheduling, FIFO scheduling and round-robin scheduling.
The obtained upper and lower bounds are useful to evaluate the delay performance of the large wireless networks. The numerical analysis shows that those bounds are tight especially for the case of heavy traffic and small active probability.

The results reveal that the offloading policy based on cell range expansion greatly reduces the macrocell traffic while bring a small amount of growth for the picocell traffic, implying that the offloading policy is potential for better utilization of the idle resources of the small cells. Our results also show that the delay performance of round-robin scheduling outperforms FIFO scheduling when the traffic is heavy, and it is reversed for light traffic since
applying round-robin scheduling may waste some time slots to schedule the empty
queues for light traffic, while in the FIFO scheduling, the scheduling may be blocked
at the users with poor channel quality for heavy traffic, resulting in large delay.

\begin{appendices}
\section{Proof of Lemma \ref{lem:succprob}}
\label{appendix:a}
\begin{proof}
The success probability conditioned on $\Phi_b$ when the typical user is served by a BS $y_0$ from the $k$th tier with link distance $Z_k=|y_0|$ is
\begin{eqnarray}
\!\!&\!\!\!\!&\!\!\mathbb{P}(\mathrm{SIR}>\theta\mid\Phi_b, y_0\in\Phi_k, Z_k=|y_0|)\nonumber\\
\!\!&\!\!=\!\!&\!\! \mathbb{P}\bigg(\frac{P_kh_{y_0}|y_0|^{-\alpha}}{\sum_{i=1}^K\sum_{y\in\Phi_i\setminus B(o,z_{ik})}\zeta_{y}P_ih_y|y|^{-\alpha}} \nonumber \\
\!\!&\!\!\!\!&\!\!>\theta\mid\Phi_b, y_0\in\Phi_k, Z_k=|y_0|\bigg)\nonumber\\
\!\!&\!\!=\!\!&\!\!\mathbb{E}\bigg[\exp\bigg(-\frac{\theta|y_0|^{\alpha}}{P_k}{\sum_{i=1}^K\sum_{y\in\Phi_i\setminus B(o,z_{ik})}\zeta_{y}P_ih_y|y|^{-\alpha}}\bigg) \nonumber\\
\!\!&\!\!\!\!&\!\!\mid\Phi_b, y_0\in\Phi_k, Z_k=|y_0|\bigg]\nonumber\\
\!\!&\!\!=\!\!&\!\!\prod_{i=1}^K\prod_{y\in\Phi_i\setminus B(o,z_{ik})}\bigg(\frac{q}{1+\theta|y_0|^{\alpha}|y|^{-\alpha}P_i/P_k}+1-q\bigg). \qquad
\end{eqnarray}

The moment generating function of $Y\stackrel{\Delta}{=}\ln\left(\mathbb{P}(\mathrm{SIR}>\theta\mid\Phi_b, y_0\in\Phi_k, Z_k=|y_0|)\right)$ is
\begin{eqnarray}
M_Y(s)\!\!&\!\!=\!\!&\!\!\mathbb{E}\left[\exp\left(sY\right)\right] \nonumber\\
\!\!&\!\!=\!\!&\!\!\mathbb{E}\left[\left(\mathbb{P}(\mathrm{SIR}>\theta\mid\Phi_b, y_0\in\Phi_k, Z_k=|y_0|)\right)^s\right] \nonumber\\
\!\!&\!\!\stackrel{(a)}{=}\!\!&\!\!\prod_{i=1}^K\mathbb{E}_{\Phi_i}\bigg[\prod_{y\in\Phi_i\setminus B(o,z_{ik})} \nonumber\\ \!\!&\!\!\!\!&\!\!\left(\frac{q}{1+\theta|y_0|^{\alpha}|y|^{-\alpha}P_i/P_k}+1-q\right)^s\bigg] \nonumber\\
\!\!&\!\!\stackrel{(b)}{=}\!\!&\!\!\prod_{i=1}^K\exp\Big(-2\pi\lambda_i\int_{z_{ik}}^\infty\Big(1-\nonumber\\
\!\!&\!\!\!\!&\!\!\Big(\frac{q}{1+\theta|y_0|^{\alpha}r^{-\alpha}P_i/P_k}+1-q\Big)^{s}\Big)r\mathrm{d}r\Big),
\end{eqnarray}
where $(a)$ follows from the independence between different tiers of BSs, and $(b)$ follows from the probability generating functional (PGFL) of the PPP. Then, we have
\begin{eqnarray}
M_Y(s)\!\!&\!\!=\!\!&\!\!\exp\Big(-2\pi\sum_{i=1}^K\lambda_i\int_{z_{ik}}^\infty\Big(1-\nonumber\\
\!\!&\!\!\!\!&\!\!\Big(\frac{q}{1+\theta|y_0|^{\alpha}r^{-\alpha}P_i/P_k}+1-q\Big)^{s}\Big)r\mathrm{d}r\Big)\nonumber\\
\!\!&\!\!=\!\!&\!\!\exp\Big(-2\pi\sum_{i=1}^K\lambda_i\int_{z_{ik}}^\infty \sum_{n=1}^\infty \binom{s}{n} (-1)^{n+1} \nonumber\\
\!\!&\!\!\!\!&\!\!\left(\frac{q\theta|y_0|^\alpha  r^{-\alpha} P_i/P_k }{1+\theta|y_0|^\alpha r^{-\alpha} P_i/P_k }\right)^n r\mathrm{d}r\Big)\nonumber\\
\!\!&\!\!=\!\!&\!\!\exp\Big(-\pi\delta\sum_{i=1}^K\lambda_i\sum_{n=1}^\infty \binom{s}{n} (-1)^{n+1} \nonumber\\
\!\!&\!\!\!\!&\!\!(q\theta|y_0|^\alpha P_i/P_k)^n \int_{z_{ik}^\alpha}^\infty \frac{u^{\delta-1}}{\left(u+\theta|y_0|^\alpha P_i/P_k\right)^n }  \mathrm{d}u\Big)\nonumber\\
\!\!&\!\!=\!\!&\!\!\exp\bigg(-\pi\delta\sum_{i=1}^K\lambda_i\sum_{n=1}^\infty \binom{s}{n} (-1)^{n+1} \Big(q\theta|y_0|^\alpha \frac{P_i}{P_k}\Big)^n \nonumber\\
\!\!\!\!\!\!\!\!\!\!&\!\!\!\!\!\!\!\!\!\!\!\!\!\!\!\!\!\!\!\!&\!\!\!\!\!\!\!\!\!\!\times\frac{z_{ik}^{\alpha(\delta-n)}}{n-\delta}{_2F_1\Big(n,n-\delta;n-\delta+1;-\frac{\theta|y_0|^\alpha P_i}{z_{ik}^\alpha P_k}\Big)}\bigg).\quad \label{equ:log_char1}
\end{eqnarray}

The cdf of $Y$, denoted by $F_Y(y)=\mathbb{P}\left(Y\leq y\right)$, follows from the Gil-Pelaez Theorem \cite{wendel1961non} as
\begin{eqnarray}
F_Y(y)=\frac{1}{2}-\frac{1}{\pi}\int_0^{\infty}\frac{\mathrm{Im}\{e^{-j\omega y}M_Y(j\omega)\}}{\omega}\mathrm{d}\omega. \label{eqn:gilpel}
\end{eqnarray}

The cdf of $\mathbb{P}(\mathrm{SIR}>\theta\mid\Phi_b, y_0\in\Phi_k, Z_k=|y_0|)$ is evaluated as
\begin{eqnarray}
&&\mathbb{P}\left(\mathbb{P}(\mathrm{SIR}>\theta\mid\Phi_b, y_0\in\Phi_k, Z_k=|y_0|)\leq u\right)\nonumber\\
&=&\mathbb{P}\left(\ln(\mathbb{P}(\mathrm{SIR}>\theta\mid\Phi_b, y_0\in\Phi_k, Z_k=|y_0|))\leq \ln u\right)\nonumber\\
&=&F_Y({\ln u})\nonumber\\
&=&\frac{1}{2}-\frac{1}{\pi}\int_0^{\infty}\frac{1}{\omega}\mathrm{Im}\bigg\{\exp\bigg(-j\omega {\ln u}\nonumber\\
&&-\pi\delta\sum_{i=1}^K\lambda_i\sum_{n=1}^\infty \binom{j\omega}{n} (-1)^{n+1} \Big(q\theta|y_0|^\alpha \frac{P_i}{P_k}\Big)^n \frac{z_{ik}^{\alpha(\delta-n)}}{n-\delta} \nonumber\\
&&\times{_2F_1\Big(n,n-\delta;n-\delta+1;-\frac{\theta|y_0|^\alpha P_i}{z_{ik}^\alpha P_k}\Big)}\bigg)\bigg\}\mathrm{d}\omega. \label{eqn:cdf_Y}
\end{eqnarray}
Plugging in $z_{ik}=\big(\frac{P_iB_i}{P_kB_k}\big)^{1/\alpha}|y_0|$, we obtain
\begin{eqnarray}
&&\mathbb{P}\left(\mathbb{P}(\mathrm{SIR}>\theta\mid\Phi_b, y_0\in\Phi_k, Z_k=|y_0|)\leq u\right)\nonumber\\
&=&\frac{1}{2}-\frac{1}{\pi}\int_0^{\infty}\frac{1}{\omega}\mathrm{Im}\bigg\{\exp\bigg(-j\omega {\ln u} \nonumber\\
&&+\pi\delta\sum_{i=1}^K\lambda_i\sum_{n=1}^\infty \binom{j\omega}{n} \Big(\frac{B_i}{B_k}\Big)^{\delta-n} \Big(\frac{P_i}{P_k}\Big)^\delta |y_0|^2   \nonumber\\
&&\times  \frac{(-q\theta)^n}{n-\delta}{_2F_1\Big(n,n-\delta;n-\delta+1;-\theta\frac{B_k}{B_i}\Big)}\bigg)\bigg\}\mathrm{d}\omega.\qquad \label{eqn:cdf_Y2}
\end{eqnarray}

The cdf of $\mathbb{P}(\mathrm{SIR}>\theta\mid\Phi_b, y_0\in\Phi_k)$ is obtained by the total probability formula as
\begin{eqnarray}
&&\mathbb{P}\left(\mathbb{P}(\mathrm{SIR}>\theta\mid\Phi_b, y_0\in\Phi_k)\leq u\right)\nonumber\\
&=& \int_0^\infty\mathbb{P}\left(\mathbb{P}(\mathrm{SIR}>\theta\mid\Phi_b, y_0\in\Phi_k, Z_k=|y_0|)\leq u\right) \nonumber\\
&&\times f_{Z_k\mid y_0\in\Phi_k}(r)\mathrm{d}r \nonumber\\
&=&\frac{2\pi \lambda_k}{\mathcal{P}_k}  \int_0^\infty\mathbb{P}\left(\mathbb{P}(\mathrm{SIR}>\theta\mid\Phi_b, y_0\in\Phi_k, Z_k=r)\leq u\right)\nonumber\\
&&\times e^{-{\pi r ^2\lambda_k}/{\mathcal{P}_k}}r\mathrm{d}r.\quad
\end{eqnarray}
Plugging in (\ref{eqn:cdf_Y2}) and calculating the integral with respective to $r$, we obtain the theorem.
\end{proof}

\section{Proof of Lemma \ref{thm:rsdelay}}
\label{appendix:b}
\begin{proof}
By the total probability formula, we obtain
\begin{eqnarray}
\mathbb{P}\left(\mathbf{D}_{\Phi_b}\leq T\right)\!\!&\!\!=\!\!&\!\!\sum_{k=1}^K \mathcal{P}_k \mathbb{P}\left(\mathbf{D}_{\Phi_b}\leq T\mid y_0\in\Phi_k\right).
\end{eqnarray}
Plugging in (\ref{eqn:randommu}) and (\ref{eqn:randomdelay}), we get the cdf as
\begin{eqnarray}
\mathbb{P}\left(\mathbf{D}_{\Phi_b}\leq T\right)\!\!&\!\!=\!\!&\!\!\sum_{k=1}^K \mathcal{P}_k \mathbb{E}_{N_k,\xi_0}\left[ \frac{1-\xi_0}{\mu-\xi_0}\leq T \mid y_0\in\Phi_k \right] \nonumber\\
\!\!&\!\!=\!\!&\!\!\sum_{k=1}^K \mathcal{P}_k \mathbb{E}_{N_k,\xi_0}\Big[\mathbb{P}\Big(\mathbb{P}(\mathrm{SIR}>\theta\mid\Phi_b, y_0\in\Phi_k) \nonumber\\
\!\!&\!\!\!\!&\!\!\geq \frac{N_k}{p}\Big(\xi_0+\frac{1-\xi_0}{T}\Big) \Big)\Big].
\end{eqnarray}
From Lemma \ref{lem:succprob}, we have
\begin{multline}
\mathbb{P}\left(\mathbf{D}_{\Phi_b}\leq T\right)=\frac{1}{2}-\frac{1}{\pi} \sum_{k=1}^K \Big[ \int_0^{\infty}\frac{1}{\omega}\mathrm{Im}\Big\{p^{j\omega}\\
\mathbb{E}_{N_k,\xi_0}\Big[\Big({N_k\Big(\xi_0+\frac{1-\xi_0}{T}\Big)}\Big)^{-j\omega}\Big]
\kappa^{-1}(\omega,k,q)
\Big\}
\mathrm{d}\omega\Big].
\end{multline}
Therefore, we obtain the results.
\end{proof}
\end{appendices}

\bibliographystyle{IEEEtran}
\bibliography{123}

\begin{IEEEbiography}[{\includegraphics[width=1in,height=1.25in,clip,keepaspectratio]{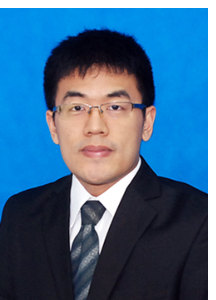}}]{Yi Zhong}
received his B.S. and Ph.D. degree in Electronic Engineering from University of Science and Technology of China (USTC) in 2010 and 2015 respectively.
From August to December 2012, he was a visiting student in Prof. Martin Haenggi's group at University of Notre Dame.
From July to October 2013, he was an research intern with Qualcomm Incorporated, Corporate Research and Development, Beijing.
From July 2015 to December 2016, he was a Postdoctoral Research Fellow with the Singapore University of Technology and Design (SUTD) in the Wireless Networks and Decision Systems (WNDS) Group led by Prof. Tony Q.S. Quek. Now, he is an assistant professor with School of Electronic Information and Communications, Huazhong University of Science and Technology, Wuhan, China.
His main research interests include heterogeneous and femtocell-overlaid cellular networks, wireless ad hoc networks, stochastic geometry and point process theory.
\end{IEEEbiography}

\begin{IEEEbiography}[{\includegraphics[width=1in,height=1.25in,keepaspectratio]{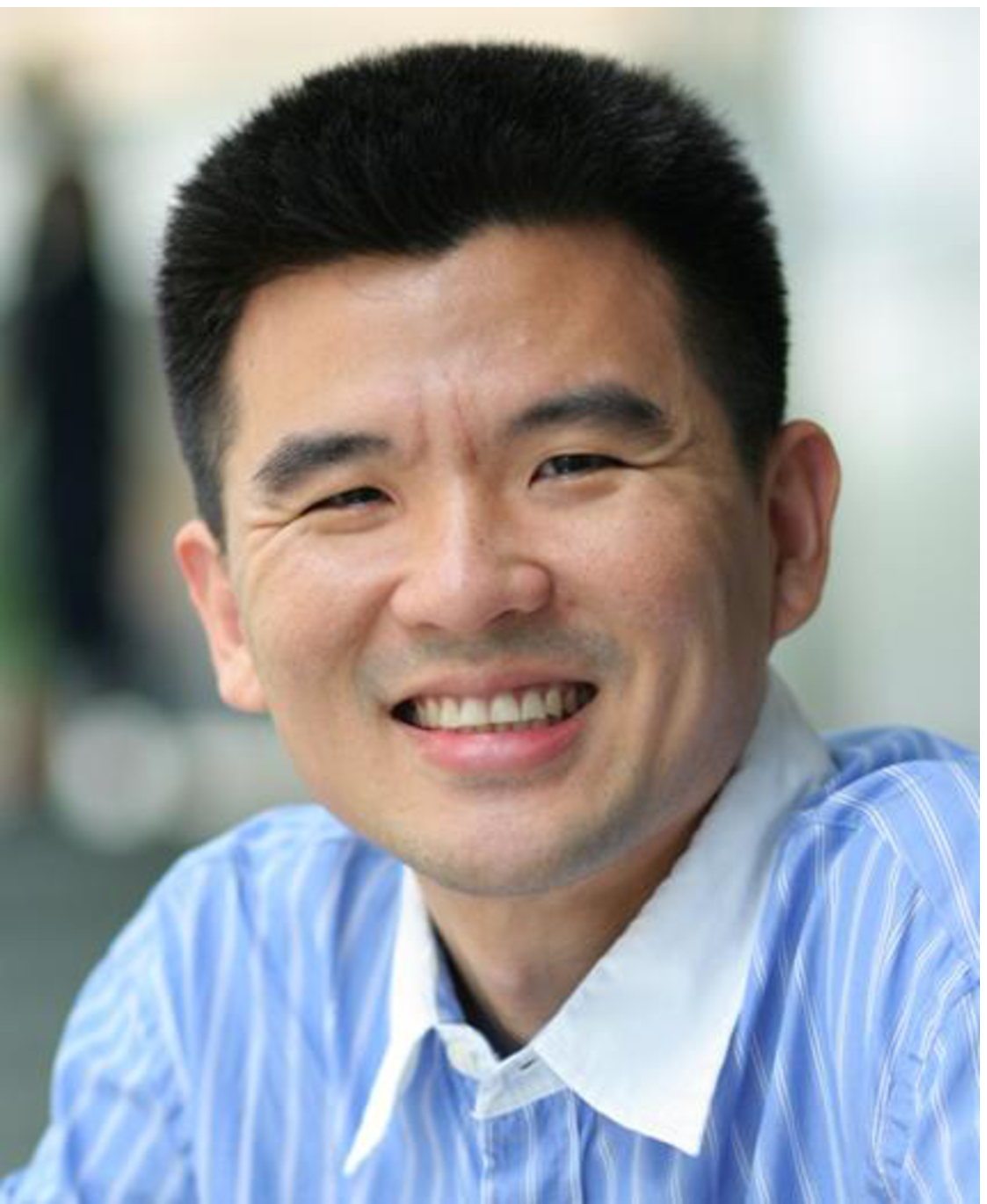}}]
{Tony Q.S. Quek}(S'98-M'08-SM'12) received the B.E.\ and M.E.\ degrees in Electrical and Electronics Engineering from Tokyo Institute of Technology, respectively. At MIT, he earned the Ph.D.\ in Electrical Engineering and Computer Science. Currently, he is a tenured Associate Professor with the Singapore University of Technology and Design (SUTD). He also serves as the Associate Head of ISTD Pillar and the Deputy Director of the SUTD-ZJU IDEA. His main research interests are the application of mathematical, optimization, and statistical theories to communication, networking, signal processing, and resource allocation problems. Specific current research topics include heterogeneous networks, green communications, wireless security, internet-of-things, and big data processing.

Dr.\ Quek has been actively involved in organizing and chairing sessions, and has served as a member of the Technical Program Committee as well as symposium chairs in a number of international conferences. He is serving as the Workshop Chair for IEEE Globecom in 2017, the Tutorial Chair for the IEEE ICCC in 2017, and the Special Session Chair for IEEE SPAWC in 2017. He is currently an Editor for the {\scshape IEEE Transactions on Communications} and an elected member of IEEE Signal Processing Society SPCOM Technical Committee. He was an Executive Editorial Committee Member for the {\scshape IEEE Transactions on Wireless Communications} and an Editor for the {\scshape IEEE Wireless Communications Letters}. He is a co-author of the book ``Small Cell Networks: Deployment, PHY Techniques, and Resource Allocation" published by Cambridge University Press in 2013 and the book ``Cloud Radio Access Networks: Principles, Technologies, and Applications" by Cambridge University Press in 2017.

Dr.\ Quek was honored with the 2008 Philip Yeo Prize for Outstanding Achievement in Research, the IEEE Globecom 2010 Best Paper Award, the 2012 IEEE William R. Bennett Prize, the IEEE SPAWC 2013 Best Student Paper Award, the IEEE WCSP 2014 Best Paper Award, the 2015 SUTD Outstanding Education Awards -- Excellence in Research, the 2016 Thomson Reuters Highly Cited Researcher, and the 2016 IEEE Signal Processing Society Young Author Best Paper Award.
\end{IEEEbiography}

\begin{IEEEbiography}[{\includegraphics[width=1in,height=1.25in,clip,keepaspectratio]{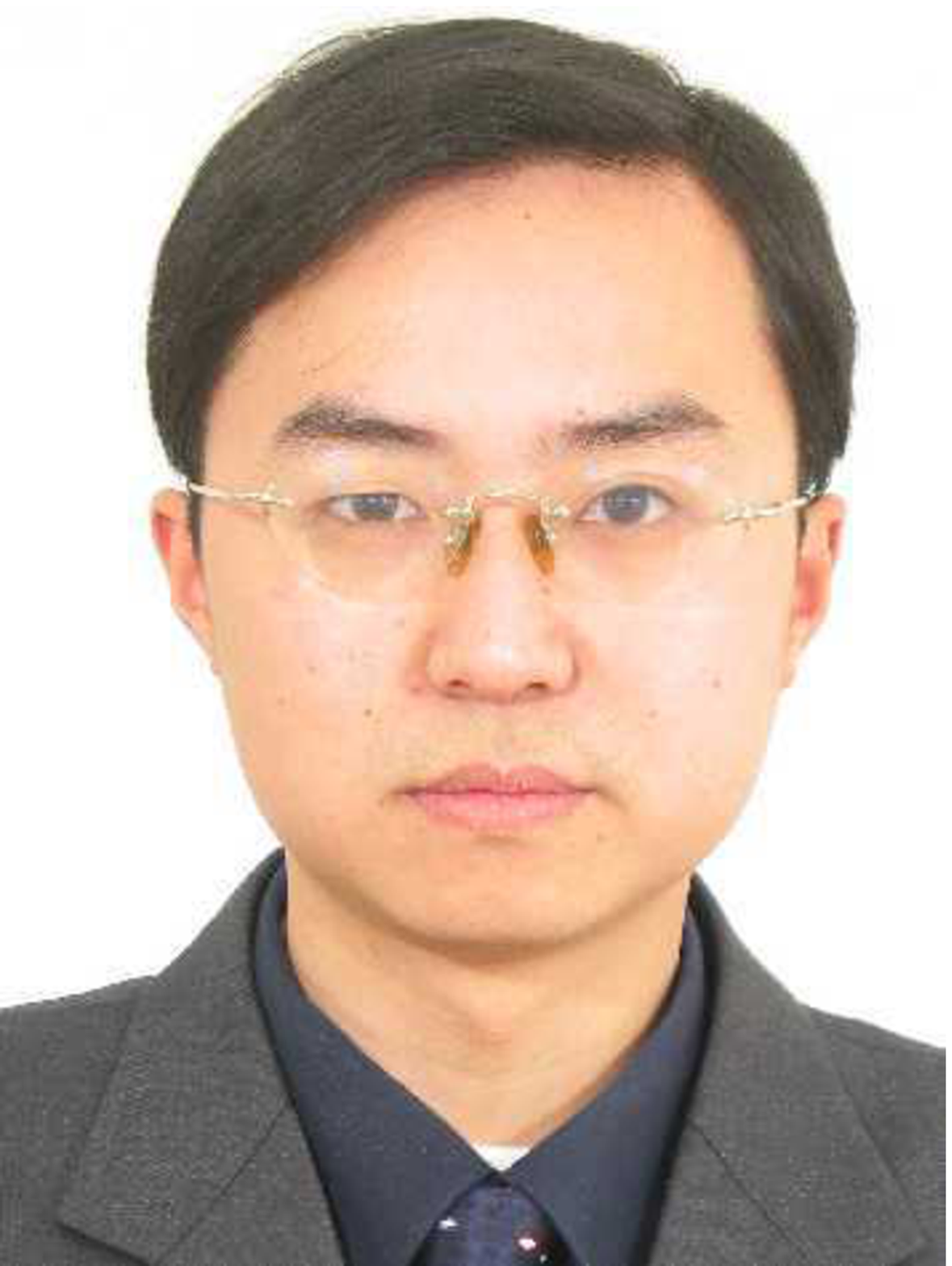}}]{Xiaohu Ge}
(M'09-SM'11) is currently a full Professor with the School of Electronic Information and Communications at Huazhong University of Science and Technology (HUST), China. He is an adjunct professor with with the Faculty of Engineering and Information
Technology at University of Technology Sydney (UTS), Australia. He is the director of China International Joint Research Center of Green Communications and Networking. He received his PhD degree in Communication and Information Engineering from HUST in 2003. He has worked at HUST since Nov. 2005. Prior to that, he worked as a researcher at Ajou University (Korea) and Politecnico Di Torino from Jan. 2004 to Oct. 2005. His research interests are in the area of mobile communications, traffic modeling in wireless networks, green communications, and interference modeling in wireless communications. He has published more than 130 papers in refereed journals and conference proceedings and has been granted about 15 patents in China. He received the Best Paper Awards from IEEE Globecom 2010.

Dr. Ge is a Senior Member of the China Institute of Communications and a member of the National Natural Science Foundation of China and the Chinese Ministry of Science and Technology Peer Review College. He has been actively involved in organizing more the ten international conferences since 2005. He served as the general Chair for the 2015 IEEE International Conference on Green Computing and Communications (IEEE GreenCom 2015). He serves as an Associate Editor for the \textit{IEEE Trancation on Green Communications and Networking}, etc.
\end{IEEEbiography}
\end{document}